\documentclass[aps,prl,showpacs,superscriptaddress,twocolumn,preprintnumbers,amsmath,amssymb]{revtex4-1}

\usepackage{array, booktabs}
\usepackage[dvipdfmx]{graphicx}
\usepackage{dcolumn}
\usepackage{bm}

\begin{document}


\title{Cavity optomagnonics with spin-orbit coupled photons}

\author{A.~Osada}
\email{alto@qc.rcast.u-tokyo.ac.jp}
\author{R.~Hisatomi}
\author{A.~Noguchi}
\author{Y.~Tabuchi}
\author{R.~Yamazaki}
\author{K.~Usami}
\email{usami@qc.rcast.u-tokyo.ac.jp}
\affiliation{Research Center for Advanced Science and Technology (RCAST), The University of Tokyo, Meguro-ku, Tokyo 153-8904, Japan}
\author{M.~Sadgrove}
\email{current affiliation: Research Institute of Electrical Communication, Tohoku University, 2-1-1 Katahira, Aoba-ku, Sendai 980-8577, Japan}
\author{R.~Yalla}
\affiliation{Center for Photonic Innovation, University of Electro-Communication, Chofu, Tokyo 182-8285, Japan}
\author{M.~Nomura}
\affiliation{Institute of Industrial Science (IIS), The University of Tokyo, Meguro-ku, Tokyo 153-8505, Japan}
\author{Y.~Nakamura$^{1, }$}
\affiliation{Center for Emergent Matter Science (CEMS), RIKEN, Wako, Saitama 351-0198, Japan}

\date{\today}

\begin{abstract}

We experimentally implement a system of \textit{cavity optomagnonics}, where a sphere of ferromagnetic material supports whispering gallery modes (WGMs) for photons and the magnetostatic mode for magnons. We observe pronounced nonreciprocity and asymmetry in the sideband signals generated by the magnon-induced Brillouin scattering of light.  The spin-orbit coupled nature of the WGM photons, their geometrical birefringence and the time-reversal symmetry breaking in the magnon dynamics impose the angular-momentum selection rules in the scattering process and account for the observed phenomena. The unique features of the system may find interesting applications at the crossroad between quantum optics and spintronics.

\begin{description}
\item[PACS numbers]
\end{description}
\end{abstract}

\maketitle



%
Spin-orbit coupling of electrons is responsible for many phenomena in condensed matter physics, such as spin-orbit splitting of the band structure~\cite{Herman1}, the spin-Hall effect~\cite{Murakami} and topological insulators~\cite{Hasan}. 
Photons also have the angular and polarization (spin) degrees of freedom, and in most cases they can be treated independently. 
However, the approximation breaks down when the spatial structure of the light mode becomes comparable to the wavelength.  
In such a case, inseparability of the orbital and spin degrees of freedom, or the spin-orbit coupling of photons, manifests itself.
While this was pointed out in the literature such as Ref.~\cite{CohenTannoudji}, it took some time before researchers acknowledged its usefulness~\cite{Onoda, Kapitanova, KY}.  Recently, the spin-orbit coupled nature of photons was vividly demonstrated in a gold nanoparticle on an optical nanofiber~\cite{Rau2} and laser-cooled Cs atoms in the vicinity of the surface of a whispering gallery mode (WGM) resonator~\cite{Rau1}.   The distinct nature of the spin-orbit coupling associated with WGMs is that the light circulating in one direction corresponds to $\sigma^+$ polarization and the other to $\sigma^-$, with respect to the direction perpendicular to the plane of the WGM orbit~\cite{Rau1}.  WGM resonators have also been intensively studied in the regime of small mode volume and the high quality factor allowing the enhancement of nonlinear optical effects~\cite{Kipp1, Kipp2, Aspelmeyer1, Tobar1}.



Here we demonstrate intriguing properties of spin-orbit coupled photons interacting with collective spin excitations in a millimeter-scale ferromagnetic sphere.  By modifying the probability of the Brillouin scattering through WGMs, one can realize cavity-assisted manipulations of magnons, leading to a new field of \textit{cavity optomagnonics}, in much the same spirit as the cavity optomechanics~\cite{Aspelmeyer1}. An additional novel feature in cavity optomagnonics is the chirality provided by the spin dynamics in the ferromagnet. This leads not only to magnon-induced nonreciprocal Brillouin scattering, but also to creation and annihilation of magnons in a highly selective manner, with the linear-polarization input.  
Moreover, WGMs coupled via an optical nanofiber will allows us to employ the resonant structures for the enhancement of the Brillouin scattering.  The combination of these properties, which was absent in the previous works~\cite{Satoh1, Satoh2}, provide a new, complementary way to investigate the magnon-induced Brillouin scattering.
We note that recent works in similar setups also observed inelastic~\cite{Hong} and elastic~\cite{Ferguson} scattering of light by the ferromagnetic spins.  It has also been shown that magnons in ferromagnets can be coherently coupled to a microwave cavity mode~\cite{Tabuchi1} as well as a superconducting qubit~\cite{Tabuchi2} in the quantum regime. Thus, the system of cavity optomagnonics presented here can open a way to optically control these degrees of freedom in the future.  (We also note that in a separate work, the authors studied the bidirectional microwave-optical conversion using a propagating light mode and a microwave cavity mode coupled to magnons~\cite{Hisatomi}.)
\begin{figure}[t]
\includegraphics{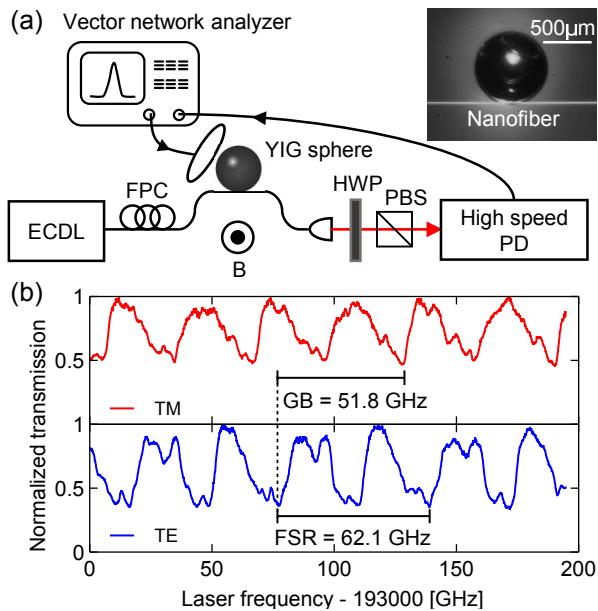}
\caption{\label{Fig1} 
Transmission through whispering gallery modes (WGMs) in a YIG sphere. (a) Experimental setup. WGMs of the YIG sphere are addressed with an optical nanofiber.  Microwave radiation from a vector network analyzer excites magnons, and AC and DC components of the light intensity are monitored with a high-speed photodetector (PD). The polarization of the light from an external-cavity diode laser (ECDL) is adjusted by a fiber polarization controller (FPC). A half- wave plate (HWP) and a polarization beam splitter (PBS) are placed before the PD. The inset shows a picture of the YIG sphere and the nanofiber. (b) Observed WGM spectra for the 750-$\mathrm{\mu}$m-diameter YIG sphere. Red and blue lines correspond to the TM and TE modes, respectively. The transmission signals are normalized by their maximal values. The free spectral range (FSR) and the estimated spectral shift due to the geometrical birefringence (GB) are indicated.
}
\end{figure}

A schematic of our experimental setup is shown in Fig.~\ref{Fig1}(a).
The WGM resonator we use is a 750-$\mathrm{\mu}$m-diameter sphere made of a ferromagnetic insulator, yttrium iron garnet (YIG). YIG is highly transparent at the optical wavelength of 1.5 $\mathrm{\mu}$m and has a refractive index of 2.19.  
With a Curie temperature of about 550 K, YIG is in the ferromagnetic phase at room temperature and supports long-wavelength magnetostatic modes~\cite{GM}.  
We focus, in particular, on the Kittel mode with spatially uniform spin precession, which exhibits a sharp ferromagnetic resonance (FMR)~\cite{Tabuchi1}.  
In order to saturate the magnetization and to define the quantization axis, a DC magnetic field $B$ of 0.24~T is applied perpendicular to the plane of the WGM orbits.  A loop coil near the YIG sphere generates an AC magnetic field perpendicular to the DC field and drives FMR.  The magnetization then acquires its horizontal component rotating at the angular frequency of the Kittel mode.  Because of the finite loss, the microwave reflection picked up by the loop coil shows a dip at the resonant frequency.  The resonant frequency and the quality factor are found to be $\omega_{\mathrm{mag}}/2\pi = 6.81$~GHz and $Q \sim 3000$, respectively.
\begin{figure}
\includegraphics{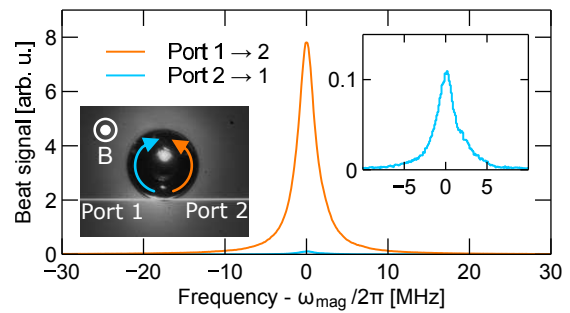}
\caption{\label{Fig2} Nonreciprocal Brillouin scattering. The orange (light-blue) plot is the observed spectrum of the beat signal for the input laser being TM-mode from Port 1 (2). The right inset shows an expanded plot of the light-blue curve. The left inset depicts the input ports and the direction of the DC magnetic field.
}
\end{figure}

Laser light with a wavelength of 1.5 $\mathrm{\mu}$m from an external cavity laser diode (ECDL) is introduced through a fiber polarization controller (FPC) and then coupled to the WGM resonator via a tapered silica optical nanofiber, with a waist diameter of about 700~nm and a waist length of around 4~mm.   Figure~\ref{Fig1}(b) shows the transmission spectra for the transverse-electric (TE) modes and the transverse-magnetic (TM) modes.  The rich structures in the spectra indicate that there are various spatial modes within the free spectral range (FSR) of 62.1~GHz.  For WGMs in the large sphere limit, frequencies of the TM modes are known to be higher than those of the TE modes with the same mode indices because of the geometrical birefringence~\cite{SB,Lam,Schiller}.  For the 750-$\mathrm{\mu}$m-diameter sphere we use, the difference is estimated to be 51.8~GHz, which is consistent with the observed spectra in Fig.~\ref{Fig1}(b) \cite{suppmat}. The intrinsic quality factors of the WGMs are found to be around $1 \times 10^5$ when they are measured in the undercoupled regime.

When the light propagates in the direction of the mean magnetization in ferromagnets, the well-known Faraday effect occurs.  When, on the other hand, the mean magnetization is perpendicular to the direction of light propagation, magnon-induced Brillouin scattering takes place~\cite{Shen1, Stancil, Wettling}.  In the presence of magnons in the Kittel mode, photons in the WGM undergo Brillouin scattering to create sideband photons with the frequency shifted by $\pm \omega_{\mathrm{mag}}/2\pi$.  A half-wave plate (HWP) and a polarization beam splitter (PBS) make the scattered sideband photons and the unscattered input photons interfere to generate a beat signal at $\omega_{\mathrm{mag}}/2\pi$. The signal is amplified and measured with a vector network analyzer.

The orange (light-blue) plot in Fig.~\ref{Fig2} shows the observed spectrum of the beat signal for the input laser being the TM-mode and coupled to the anticlockwise (clockwise) orbit of the WGM resonator. The frequency of the input photons is tuned to be $\omega/2\pi=193\,130$~GHz where the beat signal associated with the anticlockwise orbit is maximized. While both peaks in Fig.~\ref{Fig2} have the same linewidth as the FMR signal, there is a large difference in their signal strengths of almost 20~dB.

The nonreciprocity of the magnon-induced Brillouin scattering can be explained by considering the conservation of energy, momentum, and angular momentum under the situation in which the spin-orbit coupling of the photons and the geometrical birefringence associated with the WGM resonator are blended with the time-reversal symmetry breaking in the magnon dynamics. 

Suppose that the input laser polarization is adjusted to couple to the TM mode of the anticlockwise WGM orbit (orange orbit in Fig.~\ref{Fig2}). The light in the resonator is then $\sigma^+$-polarized due to the spin-orbit coupling [see Fig.~\ref{Fig3}~(a)]. To see why the Brillouin scattering is more noticeable in this situation, we consider the following three points: (\makeatletter \@roman{1} \makeatother)~The conservation of energy and spin angular momentum constrain the ensuing Brillouin scattering to create a magnon and a sideband photon at the angular frequency of $\omega - \omega_{\mathrm{mag}}$ in the $\pi$-polarized TE mode. Here we assume that the Brillouin scattering occurs only between TM and TE modes with the same WGM index, which means that the orbital angular momentum of photons is conserved. We shall return to this issue later. (\makeatletter \@roman{2} \makeatother)~Momentum conservation in the Brillouin scattering process with a magnon in the Kittel mode (having zero wavevector) leads to negligible back-scattering. This is in stark contrast with the schemes utilizing the phonons with non-zero wavevectors~\cite{Bahl, Kim, Dong, Yu} and active-passive-coupled microresonators~\cite{Peng, Chang}.  (\makeatletter \@roman{3} \makeatother)~Since the frequency of the TM mode $\omega_{\mathrm{TM}}$ is larger than that of the TE mode $\omega_{\mathrm{TE}}$ with the same mode index because of the geometrical birefringence (this is true regardless of the circulation direction of the photon~\cite{SB,Lam,Schiller}), the scattering process favors an output TE photon  with a lower frequency than that of the input TM photon, which is indeed the case for the anticlockwise orbit.

On the contrary, for the input photons in the clockwise orbit (light-blue orbit in Fig.~\ref{Fig2}) the polarizartion is $\sigma^-$ and a magnon must be annihilated. The accompanying sideband photon in the TE mode then has to have an angular frequency of  $\omega+\omega_{\mathrm{mag}}$, an unfavorable situation from the viewpoint of the geometrical birefringence, $\omega_{\mathrm{TE}}<\omega_{\mathrm{TM}}$.  This leads to the suppression of the beat signal for the clockwise photons as shown in Fig.~\ref{Fig2}. The observed small but finite signal shown in the inset can be due to the imperfection of the spin-orbit coupling associated with the WGM.

\begin{figure}[t]
\includegraphics{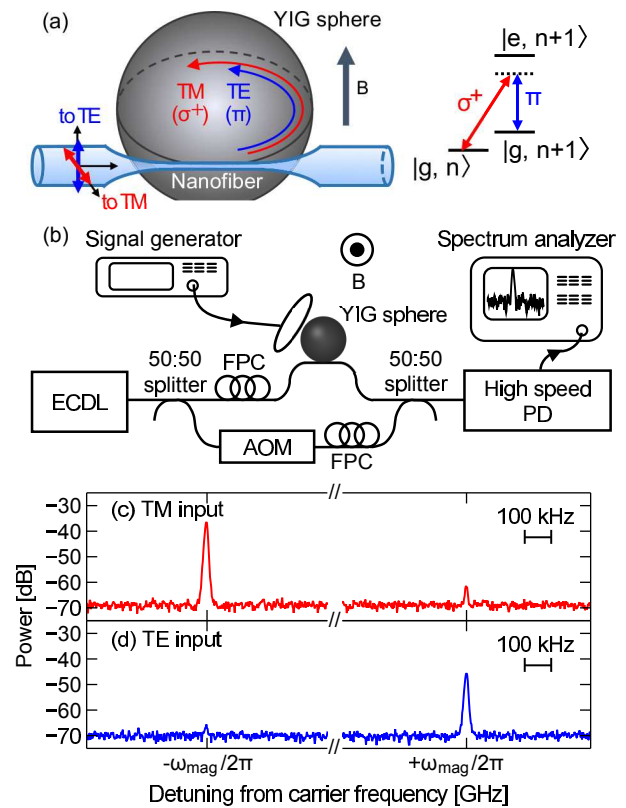}
\caption{\label{Fig3}Sideband asymmetry in magnon-induced Brillouin scattering.  (a)~Correspondence of input polarizations, polarizations in the WGM resonator, and a schematic level diagram of the Brillouin scattering. (b)~Setup for observing individual sidebands. Laser light with a wavelength of 1.5 $\mathrm{\mu}$m is split into two paths and the photons in one of the paths acts as a local oscillator (LO), whose frequency is shifted by 150~MHz by an acousto-optic modulator (AOM). For the heterodyne measurement the LO photons are mixed with the photons in the other path which are coupled to the WGM resonator via the nanofiber. Microwaves from a signal generator resonantly excite the magnons and the heterodyne signal is sent to a spectrum analyzer. (c), (d)~Observed sideband-signal powers for the TM-mode and TE-mode inputs respectively for a laser frequency of $193\,130$~GHz. The resolution bandwidth of the measurement is 10~kHz. }
\end{figure}

To clarify the selection rule in the Brillouin scattering, we consider the states $|g, n \rangle $ and $|e, n \rangle $, describing the electronic ground and excited states of the optical transition, $|g\rangle $ and $|e\rangle $, and the number of magnons in the Kittel mode, $|n\rangle$.  
If the input photons are in the TM mode the light in the resonator is $\sigma^+$-polarized as shown in Fig.~\ref{Fig3}(a) and thus the state $|g, n \rangle $ is transformed into $|g, n +1 \rangle $ via the excited state $|e, n +1 \rangle $ by creating a magnon and a down-converted red-sideband photon with $\pi$ polarization in the TE mode.
On the other hand, if the input photons are in the TE mode the light in the resonator is $\pi$-polarized and the reverse process occurs by annihilating a magnon and creating an up-converted blue-sideband photons with $\sigma^+$ polarization in the TM mode. Note that the dominant optical transition here is considered to be the spin- and parity-allowed $^6$S(3d$^5$2p$^6$) $\leftrightarrow$ $^6$P(3d$^6$2p$^5$) charge transfer transition in YIG~\cite{Shinagawa1}, whose transition wavelength is around 440 nm. Since the laser wavelength of 1.5~$\mathrm{\mu}$m is far detuned from the transition, the excited state $|e \rangle$ is only virtually populated.

To verify the asymmetry in sideband signals produced by the Brillouin scattering, the red- and blue-sideband signals are obtained in a heterodyne measurement. A schematic picture of the experimental setup is shown in Fig.~\ref{Fig3}(b).  With this scheme, the red and blue sidebands are separately observed by the spectrum analyzer at frequencies $\omega_{\mathrm{mag}}/2\pi + 150\ \mathrm{MHz}$ and $\omega_{\mathrm{mag}}/2\pi - 150\ \mathrm{MHz}$, respectively.
The results for the TM-mode input are shown in Fig.~\ref{Fig3}(c). The blue sideband is suppressed by more than 20~dB relative to the red one.  
For the TE mode input the ratio of sideband strengths is reversed as shown in Fig.~\ref{Fig3}(d).  
These results imply that by changing the polarization of the input laser, one can create or annihilate magnons in a highly controlled manner.

\begin{figure}[t]
\includegraphics{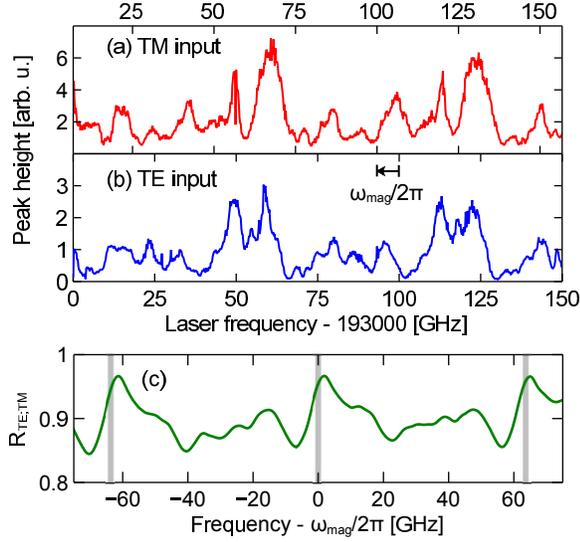}
\caption{\label{Fig4}Peak height of the beat signal vs.\ laser frequency for (a) TM-mode (with upper horizontal axis) and (b) TE-mode (with lower one) inputs.  (c) Cross correlation between the two spectra. The gray vertical lines indicate the free spectral range of the WGM.
}
\end{figure}

So far we have assumed that the Brillouin scattering occurs only between the TM and TE modes with the same WGM index and thus conserves the orbital angular-momentum of the photons. To support this hypothesis, we further analyze the Brillouin scattering strength for the TE and TM inputs.  
Let the density of states of a relevant WGM for the TM mode be $\rho^{(i)}_\mathrm{TM}(\omega)$ and that for the TE mode be $\rho^{(i)}_\mathrm{TM}(\omega)$, where $i$ represents the indices characterizing the WGMs. 
Suppose that the input photon is in the TE mode with angular frequency $\omega$ and the scattered into the TM mode with $\omega+\omega_{\mathrm{mag}}$. 
The strength of the Brillouin scattering is then written in the form proportional to the input photon number $n^{(i)}_{\mathrm{TE}}$ and then density of states of the final state $\rho^{(i)}_{\mathrm{TM}}(\omega + \omega_{\mathrm{mag}})$~\cite{suppmat}.
Since $n^{(i)}_{\mathrm{TE}}$ is proportional to $\rho^{(i)}_{\mathrm{TE}}(\omega)$, the scattering strength is written as $I_{\mathrm{TE \rightarrow TM}} (\omega) \equiv \sum_i C_i \rho^{(i)}_{\mathrm{TE}}(\omega) \rho^{(i)}_{\mathrm{TM}}(\omega + \omega_{\mathrm{mag}})$ with coefficient $C_i $ representing the contributions from each mode.  
For the TM mode input, the same consideration results in $I_{\mathrm{TM \rightarrow TE}}(\omega) \equiv \sum_i C_i \rho^{(i)}_{\mathrm{TM}}(\omega) \rho^{(i)}_{\mathrm{TE}}(\omega - \omega_{\mathrm{mag}})$.  
Thus, one can immediately see that $I_{\mathrm{TE \rightarrow TM}}(\omega) = I_{\mathrm{TM \rightarrow TE}}(\omega + \omega_{\mathrm{mag}})$.


We measure the laser-frequency dependence of the beat signal in the setup shown in Fig.~\ref{Fig1}(a) for the TM- and TE-mode inputs. The observed spectra are compared in Figs.~\ref{Fig4}(a) and (b), and indeed we find apparent similarity.
We characterize the similarity between the two spectra by calculating the cross correlation between them.
Here the cross correlation $R_{\mathrm{TE;TM}}$ is defined as
\begin{equation*}
R_{\mathrm{TE;TM}}(\Omega) = \frac{\left(\int_{0}^{\infty} \frac{\mathrm{d}\omega}{2\pi} \sqrt{ I^{\mathrm{(exp)}}_{\mathrm{TE \rightarrow TM}}(\omega) I^{\mathrm{(exp)}}_{\mathrm{TM \rightarrow TE}}(\omega + \Omega)}\right)^2}{\int_{0}^{\infty} \frac{\mathrm{d}\omega}{2\pi}  I^{\mathrm{(exp)}}_{\mathrm{TE \rightarrow TM}}(\omega)   \int_{0}^{\infty} \frac{\mathrm{d}\omega}{2\pi}  I^{\mathrm{(exp)}}_{\mathrm{TM \rightarrow TE}}(\omega + \Omega) },
\end{equation*}
where the quantities with the superscript ``(exp)" denotes the experimentally observed spectra.  The quantity $R_{\mathrm{TE;TM}}$ equals unity when $I^{\mathrm{(exp)}}_{\mathrm{TE \rightarrow TM}}(\omega) \propto I^{\mathrm{(exp)}}_{\mathrm{TM \rightarrow TE}}(\omega + \Omega)$.
The obtained cross correlation $R_{\mathrm{TE;TM}}$ is shown in Fig.~\ref{Fig4}(c).
The gray vertical lines in the figure indicate the free spectral range of the WGMs ($0$ GHz and $\pm 62.1$ GHz). 
The maxima of the cross correlation at these frequencies qualitatively supports the similarity between the spectra and underlying assumption that the Brillouin scattering process conserves the orbital angular momentum of the photons.

Note also that the two spectra in Figs.~\ref{Fig4}(a) and (b) do not match the WGM transmission spectra of corresponding polarizations in Fig.~\ref{Fig1}(b). This discards the simple proportionality between the Brillouin-scattering and the transmission spectra and fortifies the claim that the strengths of the Brillouin scattering processes are proportional to $I_{\mathrm{TE \rightarrow TM}}$ and $I_{\mathrm{TM \rightarrow TE}}$, respectively. Complete assignment of the WGMs is needed for more quantitative understanding of the spectral structure. From this perspective, a WGM resonator with a higher quality factor and a smaller number of relevant spatial modes will be of great help.

For a candidate of a microwave-to-optical-photon quantum transducer, the coupling constant $g$ of the magnon-induced Brillouin scattering and the microwave-to-optical photon conversion efficiency are crucial parameters.  The coupling constant is theoretically given by
$g^{(\mathrm{theory})} = \mathcal{V} c' \sqrt{2/N_{\rm{spin}}} = 2\pi\times 5.4 \,\rm{Hz}$ where $\mathcal{V}$, $c'$ and $N_{\rm{spin}}$ are respectively the Verdet constant, the speed of light inside the material and the number of spins in the sample~\cite{suppmat}.  The experimentally obtained coupling constant $g^{(\mathrm{exp})}$ of $2\pi\times 5 \,\rm{Hz}$ is consistent with the theoretical value~\cite{suppmat}.  Since the decay rates of the Kittel mode and the WGMs are in the MHz and GHz range, our system is in the weak coupling regime.

In our current setup, the maximum microwave-to-optical photon conversion efficiency is $7\times10^{-14}$~\cite{suppmat}, even lower than that in the experiment without a WGM resonator~\cite{Hisatomi}.  The main reason for the small value is the frequency mismatch between the frequency difference of the TE and TM WGMs and the Kittel-mode frequency, which can be solved by properly designed WGM resonator geometries.  Another reason is the modest quality factors of the WGMs, which can be improved up to the absorption-limited value of $3\times10^6$ \cite{Wood}.  With these improvements, the conversion efficiency will reach $3\times10^{-2}$~\cite{suppmat}.  Materials with larger Verdet constant allows further enhancement by orders of magnitude to make the transduction feasible in the quantum regime.

In conclusion, we observed magnon-induced nonreciprocal Brillouin scattering in a sphere of ferromagnetic insulator material. The phenomena are subject to the unique selection rule imposed by the spin-orbit coupled nature of the WGM photons, the geometrical birefringence of the WGM resonator, and the time-reversal symmetry breaking in the magnon dynamics. The selection rule is also responsible for the sideband asymmetry in the Brillouin scattering process, providing us with a powerful tool to selectively create or annihilate magnons in the Kittel mode with optical photons.     These unique features of the system allow it to serve as an interesting testbed for investigating the interdisciplinary field involving quantum optics and spintronics.

We would like to thank Kohzo Hakuta for his advice. This work was supported by the Project for Developing Innovation System of MEXT, JSPS KAKENHI (grant no. 26600071, 26220601, 15H05461), the Murata Science Foundation, the Inamori Foundation, Research Foundation for Opto-Science and Technology, and NICT.

\newpage

\newcommand{\1}{I}
\newcommand{\2}{I\hspace{-.1em}I}
\newcommand{\3}{I\hspace{-.1em}I\hspace{-.1em}I}
\newcommand{\4}{I\hspace{-.1em}V}
\newcommand{\5}{V}
\newcommand{\6}{V\hspace{-.1em}I}
\newcommand{\7}{V\hspace{-.1em}I\hspace{-.1em}I}
\newcommand{\8}{V\hspace{-.1em}I\hspace{-.1em}I\hspace{-.1em}I}
\newcommand{\9}{I\hspace{-.1em}X}

\newcommand{\atea}{a^{\hspace{0.1mm}}_{\mathrm{TE}}}
\newcommand{\atec}{a^{\dagger}_{\mathrm{TE}}}
\newcommand{\atma}{a^{\hspace{0.1mm}}_{\mathrm{TM}}}
\newcommand{\atmc}{a^{\dagger}_{\mathrm{TM}}}
\newcommand{\ba}{b}
\newcommand{\bc}{b^{\dagger}_{\hspace{0.1mm}}}
\newcommand{\Ateia}{A^{(\mathrm{in})}_{\mathrm{TE}}}
\newcommand{\Ateic}{A^{(\mathrm{in}) \dagger}_{\mathrm{TE}}}
\newcommand{\Atmia}{A^{(\mathrm{in})}_{\mathrm{TM}}}
\newcommand{\Atmic}{A^{(\mathrm{in}) \dagger}_{\mathrm{TM}}}
\newcommand{\Bia}{B^{(\mathrm{in})}_{\hspace{0.1mm}}}
\newcommand{\Bic}{B^{(\mathrm{in}) \dagger}_{\hspace{0.1mm}}}
\newcommand{\Ateoa}{A^{(\mathrm{out})}_{\mathrm{TE}}}
\newcommand{\Ateoc}{A^{(\mathrm{out}) \dagger}_{\mathrm{TE}}}
\newcommand{\Atmoa}{A^{(\mathrm{out})}_{\mathrm{TM}}}
\newcommand{\Atmoc}{A^{(\mathrm{out}) \dagger}_{\mathrm{TM}}}
\newcommand{\Boa}{B^{(\mathrm{out})}_{\hspace{0.1mm}}}
\newcommand{\Boc}{B^{(\mathrm{out}) \dagger}_{\hspace{0.1mm}}}
\newcommand{\wte}{\omega^{\hspace{0.1mm}}_{\mathrm{TE}}}
\newcommand{\wtm}{\omega^{\hspace{0.1mm}}_{\mathrm{TM}}}
\newcommand{\wm}{\omega^{\hspace{0.1mm}}_{\mathrm{m}}}
\newcommand{\Wte}{\Omega^{\hspace{0.1mm}}_{\mathrm{TE}}}
\newcommand{\Wtm}{\Omega^{\hspace{0.1mm}}_{\mathrm{TM}}}
\newcommand{\Wm}{\Omega^{\hspace{0.1mm}}_{\mathrm{m}}}
\newcommand{\gte}{\gamma^{\hspace{0.1mm}}_{\mathrm{TE}}}
\newcommand{\gtm}{\gamma^{\hspace{0.1mm}}_{\mathrm{TM}}}
\newcommand{\gm}{\gamma^{\hspace{0.1mm}}_{\mathrm{m}}}
\newcommand{\Gte}{\Gamma^{\hspace{0.1mm}}_{\mathrm{TE}}}
\newcommand{\Gtm}{\Gamma^{\hspace{0.1mm}}_{\mathrm{TM}}}
\newcommand{\Gm}{\Gamma^{\hspace{0.1mm}}_{\mathrm{m}}}
\newcommand{\kte}{\kappa^{\hspace{0.1mm}}_{\mathrm{TE}}}
\newcommand{\ktm}{\kappa^{\hspace{0.1mm}}_{\mathrm{TM}}}
\newcommand{\km}{\kappa^{\hspace{0.1mm}}_{\mathrm{m}}}
\newcommand{\Dte}{\Delta^{\hspace{0.1mm}}_{\mathrm{TE}}}
\newcommand{\Dtm}{\Delta^{\hspace{0.1mm}}_{\mathrm{TM}}}
\newcommand{\Dm}{\Delta^{\hspace{0.1mm}}_{\mathrm{m}}}
\newcommand{\Dtesq}{\Delta^{2}_{\mathrm{TE}}}
\newcommand{\Dtmsq}{\Delta^{2}_{\mathrm{TM}}}
\newcommand{\Dmsq}{\Delta^{2}_{\mathrm{m}}}
\newcommand{\alte}{\alpha^{\hspace{0.1mm}}_{\mathrm{TE}}}
\newcommand{\altm}{\alpha^{\hspace{0.1mm}}_{\mathrm{TM}}}
\newcommand{\betam}{\beta^{\hspace{0.1mm}}_{\mathrm{m}}}
\newcommand{\ntei}{n^{(\mathrm{in})}_{\mathrm{TE}}}
\newcommand{\ntmi}{n^{(\mathrm{in})}_{\mathrm{TM}}}
\newcommand{\nmi}{n^{(\mathrm{in})}_{\mathrm{MW}}}
\newcommand{\nteo}{n^{(\mathrm{out})}_{\mathrm{TE}}}
\newcommand{\ntmo}{n^{(\mathrm{out})}_{\mathrm{TM}}}
\newcommand{\nmo}{n^{(\mathrm{out})}_{\mathrm{MW}}}
\newcommand{\rte}{\rho^{\hspace{0.1mm}}_{\mathrm{TE}}}
\newcommand{\rtm}{\rho^{\hspace{0.1mm}}_{\mathrm{TM}}}
\newcommand{\rmag}{\rho^{\hspace{0.1mm}}_{\mathrm{m}}}

\newcommand{\dt}[1]{\frac{\mathrm{d} {#1}}{\mathrm{d}t}}
\newcommand{\beginsupplement}{%
        \setcounter{table}{0}
        \renewcommand{\thetable}{S\arabic{table}}%
        \setcounter{figure}{0}
        \renewcommand{\thefigure}{S\arabic{figure}}%
     }


\title{Cavity optomagnonics with spin-orbit coupled photons: Supplemental Material}

\author{A.~Osada$^{1}$}
\email{alto@qc.rcast.u-tokyo.ac.jp}
\author{R.~Hisatomi$^{1}$}
\author{A.~Noguchi$^{1}$}
\author{Y.~Tabuchi$^{1}$}
\author{R.~Yamazaki$^{1}$}
\author{K.~Usami$^{1}$}
\email{usami@qc.rcast.u-tokyo.ac.jp}
\author{M.~Sadgrove$^{2}$}
\email{current affiliation: Research Institute of Electrical Communication, Tohoku University, 2-1-1 Katahira, Aoba-ku, Sendai 980-8577, Japan}
\author{R.~Yalla$^{2}$}
\author{M.~Nomura$^{3}$}
\author{Y.~Nakamura$^{1, 4}$ \\ \hspace{1mm}  \\}

\affiliation{$^1$Research Center for Advanced Science and Technology (RCAST), The University of Tokyo, Meguro-ku, Tokyo 153-8904, Japan}
\affiliation{$^2$Center for Photonic Innovation, University of Electro-Communication, Chofu, Tokyo 182-8285, Japan}
\affiliation{$^3$Institute of Industrial Science (IIS), The University of Tokyo, Meguro-ku, Tokyo 153-8505, Japan}
\affiliation{$^4$Center for Emergent Matter Science (CEMS), RIKEN, Wako, Saitama 351-0198, Japan}


\date{\today}


\maketitle

\beginsupplement
\begin{table*}[t]
\caption{Symbols used in the Supplemental Material}
\begin{center}
\begin{tabular}{lc}
\toprule
Quantities & Symbols \\
\midrule
Frequency of TE (TM) WGM and Kittel mode \hspace{5mm} & $\Wte$ ($\Wtm$), $\Wm$ \\
Intrinsic decay rate of TE (TM) WGM and Kittel mode \hspace{5mm} & $\gte$ ($\gtm$), $\gm$ \\
Creation and annihilation operators of TE (TM) WGM \hspace{5mm} & $\atec, \atea$ ($\atmc, \atma$) \\
Creation and annihilation operators of Kittel mode \hspace{5mm} & $\bc, \ba$ \\
Frequency of input or output field for TE (TM) WGM  and Kittel mode \hspace{5mm} & $\wte$ ($\wtm$), $\wm$\\
External coupling to TE (TM) WGM and Kittel mode \hspace{5mm} & $\kte$ ($\ktm$), $\km$\\
Creation and annihilation operators of input and output fields for TE WGM \hspace{5mm} & $\Ateic, \Ateia, \Ateoc, \Ateoa$ \\
Creation and annihilation operators of input and output fields for TM WGM \hspace{5mm} & $\Atmic, \Atmia, \Atmoc, \Atmoa$ \\
Creation and annihilation operators of input and output fields for Kittel mode \hspace{5mm} & $\Bic, \Bia, \Boc, \Boa$ \\
Intra-cavity photon numbber of TE (TM) WGM \hspace{5mm} & $n^{\hspace{1mm}}_{\rm{TE}}$ ($n^{\hspace{1mm}}_{\rm{TM}}$) \\
Input and output photon flux for TE (TM) WGM \hspace{5mm} & $\ntei, \nteo$ ($\ntmi, \ntmo$) \\
Input and output photon flux for Kittel mode \hspace{5mm} & $\nmi, \nmo$ \\
Interaction Hamiltonian for CCW and CW cases & $\mathcal{H}_{\rm{int}}^{(\mathrm{CCW})}, \mathcal{H}_{\rm{int}}^{(\mathrm{CW})}$ \\
Detuning from TE (TM) WGM resonance, $\Wte - \wte$ ($\Wtm - \wtm$) \hspace{5mm} & $\Dte$ ($\Dtm$) \\
Detuning from Kittel-mode resonance, $\Wm - \wm$ \hspace{5mm} & $\Dm$ \\
$\gte + \kte$, $\gtm + \ktm$, $\gm + \km$\hspace{5mm} & $\Gte$, $\Gtm$, $\Gm$\\
Coupling coefficient of magnon-induced Brillouin scattering process \hspace{5mm} & $g$ \\
\bottomrule
\end{tabular}
\end{center}
\end{table*}

\begin{figure*}[b]
\includegraphics{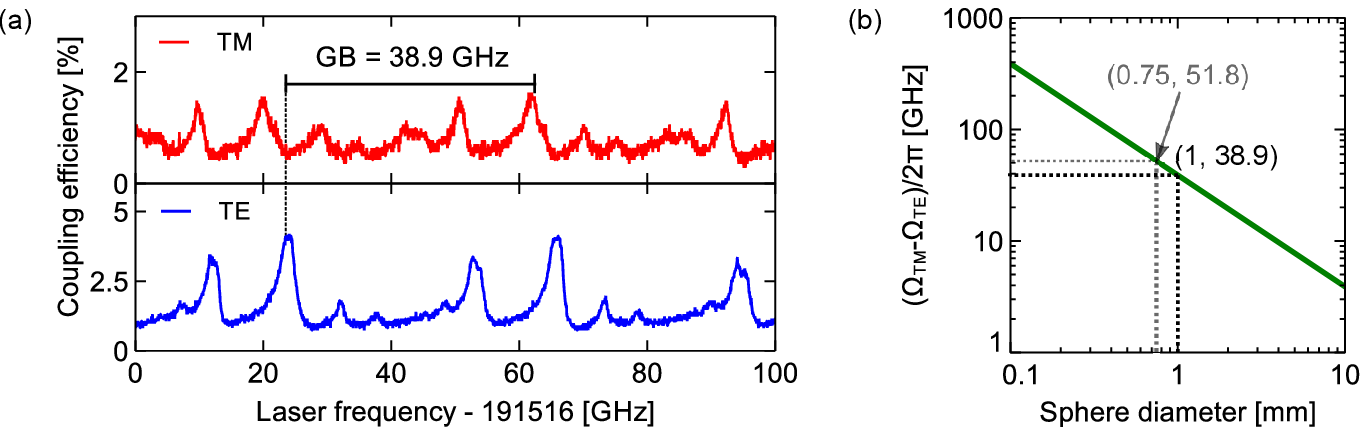}
\caption{\label{FigA5} Difference between $\Wtm$ and $\Wte$ due to the geometrical birefringence (GB).  (a) Spectra of WGMs in a 1mm-diameter YIG sphere.  The theoretically estimated value of the frequency difference $(\Wtm-\Wte)/2\pi$ due to the GB is indicated.  (b) $(\Wtm-\Wte)/2\pi$ as a function of the sphere diameter.  Predicted values of $(\Wtm-\Wte)/2\pi$ for the sphere diameters of 0.75~mm and 1~mm are explicitly shown.
}
\end{figure*}
\section{Supplement A: Estimation of the frequency difference between TE and TM WGMs}

Here we intend to discuss the consistency of the estimation and the experimentally observed whispering-gallery-mode (WGM) spectra in terms of the frequency difference between TE and TM WGMs.
Some of the symbols used here are defined in Table~S1.
From the analytical solution of  WGM resonances in a dielectric sphere in the limit of large mode's order \cite{SB,Lam,Schiller}, we can write the frequency difference between TE and TM WGMs due to the geometrical birefringence (GB) as
\begin{align}
\frac{(\Omega_{\rm{TM}}-\Omega_{\rm{TE}})}{2\pi} = \frac{c}{\pi n_r D} \sqrt{1-\frac{1}{n_r^2}}, \label{tm-te} \tag{S1}
\end{align}
where $D$ being the diameter of the resonator and $n_r=2.19$ the refractive index.  Figure.~\ref{FigA5}(a) shows the WGM spectra in a 1mm-diameter YIG sphere coupled via a prism.  A high-refractive-index ($\sim3.6$) silicon prism is used and the measurements are done in the weak coupling regime.  The vertical axis represents the coupling efficiency, the percentage of the intensity drop of light around the resonances.  The quantity $(\Omega_{\rm{TM}}-\Omega_{\rm{TE}})/2\pi$ is plotted as a function of the sphere diameter in Fig.~\ref{FigA5}(b).  The experimentally observed spectra such as the ones in Fig.~\ref{FigA5}(a) and Fig.~1(b) in the main article are consistent with the analytical solution of Eq.~(\ref{tm-te}).

\section{Supplement B: Theory of cavity optomagnonics}

\subsection{Description of the interaction Hamiltonian and the optomagnonic coupling constant}

In this section we analyze the system of cavity optomagnonics and the photon-number conversion efficiency with the use of the input-output theory.  The symbols used here are summarized in Table~S1.  The novel feature underlying the system is to make use of the polarization states of the WGMs.
The WGMs have the following important properties: first, it has its spin angular momentum perpendicular to the plane of the orbit. Second, the TM mode is \textit{spin-orbit coupled}; one direction corresponds to $\sigma^+$ polarization and the other to $\sigma^-$~\cite{Rau1, Rau2}.
The situation we consider is that this spin-orbit-coupled WGM photons interact with the Kittel-mode magnons under the external magnetic field applied perpendicular to the plane of the WGM orbit.  In such a configuration, the Kittel-mode magnon interacts with almost purely $\sigma^+$-, $\sigma^-$- or $\pi$-polarized photon depending on the polarization and the direction of the WGM orbit.  Limiting the direction to only one of the two, one may have a set of correspondence $(\mathrm{TE}, \mathrm{TM}) = (\pi, \sigma^+)$ for the counterclockwise (CCW) orbit or $(\pi, \sigma^-)$ for the clockwise (CW) one.  For example, when a CCW WGM is chosen the $\sigma^+$ and $\pi$ polarized photons are exchanged to each other by the magnon Brillouin scattering process.  Together with the spin conservation, the interaction Hamiltonians read
\begin{align}
\mathcal{H}_{\mathrm{int}}^{(\mathrm{CCW})} &= \hbar g (\atea \atmc \ba + \atec \atma \bc), \label{Hccw} \tag{S2}\\
\hspace{5mm}\mathcal{H}_{\mathrm{int}}^{(\mathrm{CW})} &= \hbar g (\atea \atmc \bc + \atec \atma \ba), \notag
\end{align}
where $g$ is the coupling constant.

The interaction Hamiltonian of Eq.~(\ref{Hccw}) can also be obtained by considering the volume integration of the interaction-energy density
\begin{align} 
-i\epsilon_0 f M_y E_{\rm{TM}}^{*} E_{\rm{TE}} \label{intMEE} \tag{S3}
\end{align}
and its complex conjugate.  Here $E_{\rm{TM}}$ and $E_{\rm{TE}}$ denote the electric fields of the TM and TE WGMs, while $M_y$ the $y$-component of the magnetization $(M_x, M_y, M_z)$ where the mean magnetization is oriented along $z$-axis.  The coefficient $f$ originates in the permittivity tensor describing the magneto-optical effect which is expressed in the Cartesian coordinates as~\cite{Stancil}
\begin{align}
\epsilon = \epsilon_0 \left( \begin{matrix}
		\epsilon_r 		& -ifM_z 			& ifM_y \\
		ifM_z 		& \epsilon_r 		& -ifM_x \\
		-ifM_y		& ifM_x			& \epsilon_r
		\end{matrix} \right) \tag{S4}
\end{align}
with the relative permittivity $\epsilon_r$.  In a saturated ferromagnet the coefficient $f$ is related to the Verdet constant $\mathcal{V}$ as $f = (2\sqrt{\epsilon_r}/k_0 M_z) \mathcal{V}$~\cite{Stancil} with the wavevector $k_0$ of the optical field in the vacuum.  Note here that whether $M_x$ or $M_y$ appears in the relevant interaction term Eq.~(\ref{intMEE}) depends on the choice of the coordinates.  The three quantities, $M_y$, $E_{\rm{TM}}$ and $E_{\rm{TE}}$, are now translated into the operators $\ba$, $\atma$ and $\atea$ with some coefficients to give interaction terms $\hbar g \atea \atmc \ba$ and its Hermitian conjugate.  The pre-factors for the operators $\ba$, $\atma$ and $\atea$ are written respectively by $\sqrt{2g_s\mu_{\rm{B}} M_z/V}$, $\sqrt{\hbar\Wtm/2\epsilon_0\epsilon_r V_{\rm{TM}}}$ and $\sqrt{\hbar\Wte/2\epsilon_0\epsilon_r V_{\rm{TE}}}$, where the sample volume $V$, the mode volumes $V_{\rm{TM}}$ and $V_{\rm{TE}}$ of TM and TE WGMs, the electronic $g$-factor $g_s$ and the Bohr magneton $\mu_{\rm{B}}$ are used.  Suppose $\Wtm \simeq \Wte \equiv \Omega$ and $V_{\rm{TM}} = V_{\rm{TE}} \equiv V_{\rm{WGM}}$, and the volume integration runs over the interaction region to give a factor of $V_{\rm{WGM}}$.  Thus, we get the theoretical expression $g^{(\rm{theory})}$ for the coupling strength
\begin{align} 
&g^{(\rm{theory})} \notag\\
&\hspace{5mm} = \frac{1}{\hbar} \epsilon_0 f V_{\rm{WGM}} \sqrt{\frac{2g_s\mu_{\rm{B}} M_z}{V}} \sqrt{\frac{\hbar\Wtm}{2\epsilon_0\epsilon_r V_{\rm{TM}}}} \sqrt{\frac{\hbar\Wte}{2\epsilon_0\epsilon_r V_{\rm{TE}}}} \notag \\
&\hspace{5mm} = \mathcal{V} \frac{c}{n_r} \sqrt{\frac{2}{n_{\mathrm{spin}} V}} \label{gth} \tag{S5}
\end{align}
where $c$ denotes the speed of light and $n_{\mathrm{spin}}$ the spin density. In the above expression we used the relation $M_z = g_s \mu_B n_{\rm{spin}}$.  Alternatively, with the speed of light inside the material $c'$ and the number of spins $N_{\mathrm{spin}} = n_{\mathrm{spin}} V$ we can also write this in the suggestive form:
\begin{align} 
g^{(\rm{theory})} = \mathcal{V} c' \sqrt{\frac{2}{N_{\mathrm{spin}}}}.   \tag{S6}
\end{align}
This expression implies that the optomagnonical coupling is proportional to the Verdet constant $\mathcal{V}$ and furthermore the spin fluctuation appears in the form of $\sqrt{1/N_{\mathrm{spin}}}$, as we expect.  This quantity can also be regarded as the Faraday rotation angle per unit time due to the vacuum fluctuation of the Kittel mode.

In the case of our experiment, with the Verdet constant $\mathcal{V} = 3.77$~$\mathrm{rad/cm}$, the spin density $n_{\mathrm{spin}} = 2.1 \times 10^{28}$~$/\mathrm{m}^3$, the sample volume $V = (4\pi/3)\times(0.375)^3$~$\mathrm{mm}^3$, and the refractive index $n_r = 2.19$, the theoretical coupling strength $g^{(\rm{theory})}$ is evaluated to be $2\pi \times$5.4~Hz.  

\subsection{Input-output formalism of the cavity optomagnonical system}

\begin{figure*}[t]
\includegraphics{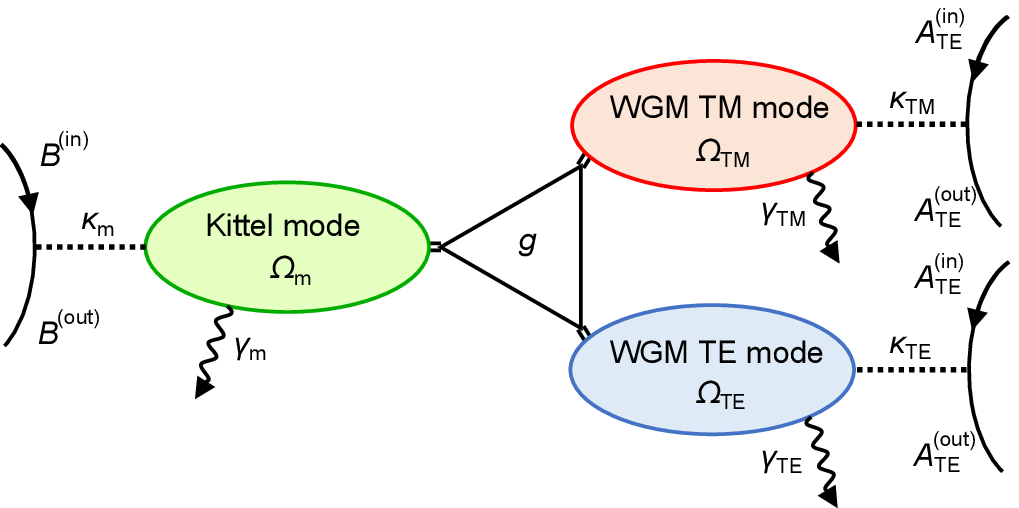}
\caption{\label{FigA2} Schematic diagram of the system with the parameters used in the calculation.  The TE and TM WGMs and the Kittel mode are connected by the three-wave process depicted by the triangle at the center.
}
\end{figure*}

\begin{figure}[t]
\includegraphics{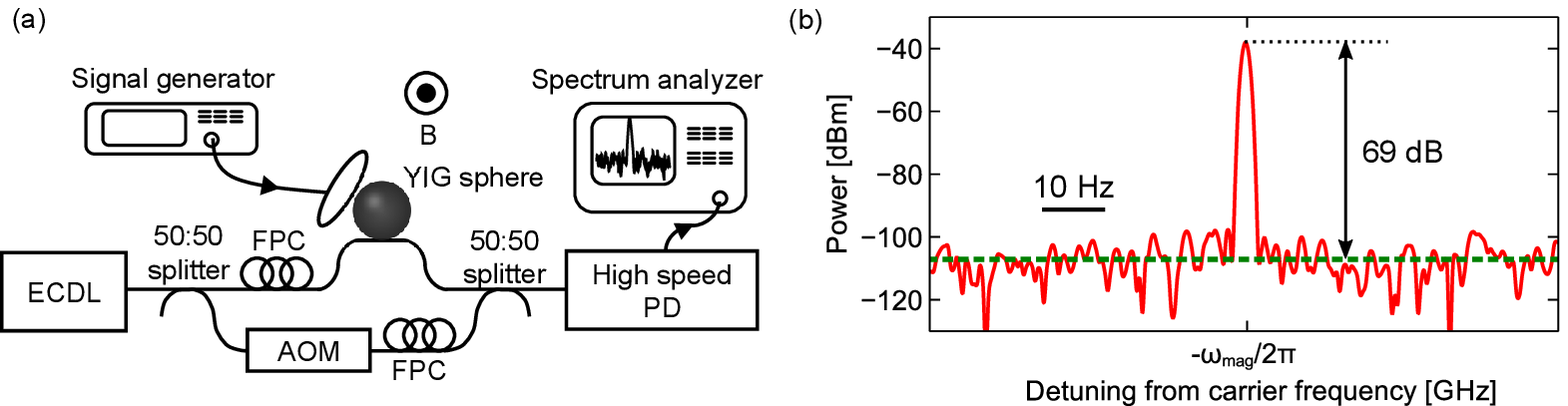}
\caption{\label{FigB1} Estimation of the conversion efficiency by the measurement of the strength of the sideband signal and its ratio to the shot-noise power level.  (a) Experimental setup.  (b) Sideband power for the TM polarization measured by the spectrum analyzer with the resolution bandwidth of 1 Hz.
}
\end{figure}

A schematic diagram of the whole optomagnonical system is shown in Fig.~\ref{FigA2}.  Below we focus on the case of CCW WGMs, namely the photons in the TM mode can be regarded as $\sigma^+$-polarized photons.  Then the Hamiltonian $\mathcal{H}_0$ of the whole system is given as the sum of the energies of the TE and TM WGMs and the Kittel mode, and the interaction Hamiltonian $\mathcal{H}_{\mathrm{int}}^{(\mathrm{CCW})}$ in Eq.~(\ref{Hccw}). 
Suppose that the TM mode is driven by a laser with the frequency $\wtm$.  This allows us to replace $\atma$ with a classical field $\sqrt{n^{\hspace{0.1mm}}_{\mathrm{TM}}} e^{ -i\wtm t }$, where $n^{\hspace{0.1mm}}_{\rm{TM}} = \ntmi \ktm/\left[\Dtmsq + \left( \Gtm /2 \right)^2 \right] $ denotes the average number of photons in the TM WGM.  Now we move to a rotating frame by performing the unitary transformation $U(t) =$ $\exp{\left[i\wtm \atmc \atma t+i\wte \atec \atea t+i\wm \bc \ba\, t\right]}$ which transforms the Hamiltonian in the form
\begin{align}
\mathcal{H} &= U^{\dagger}(t) \mathcal{H} U(t) \notag\\
&=  \hbar \Dte \atec \atea + \hbar \Dtm \atmc \atma + \hbar \Dm \bc \ba  \notag \\
&\hspace{5mm}+ \hbar g \sqrt{n^{\hspace{0.1mm}}_{\mathrm{TM}}} (\atea \ba\, e^{i\delta t} + \atec \bc e^{-i\delta t}). \label{H} \tag{S7}
\end{align}
There appears the factor $e^{i\delta t}$, which imposes the frequency matching condition $\delta \equiv \wtm - \wte - \wm =0$.  As a result we obtain a set of equations of motion:
\begin{align}
\dt{\atec} &=  i \Dte \atec + i g \sqrt{n^{\hspace{0.1mm}}_{\mathrm{TM}}} \ba  - \frac{\Gte}{2} \atec - \sqrt{\kte} \Ateia (t), \label{eqatea} \tag{S8}\\
\dt{\ba} &=   -i \Dm \ba - i g \sqrt{n^{\hspace{0.1mm}}_{\mathrm{TM}}} \atec - \frac{\Gm}{2}\ba - \sqrt{\km} \Bia (t). \label{eqba} \tag{S9}
\end{align}
We can adiabatically eliminate the TE WGM, assuming that $\Gte$ is the largest value in the situation considered.  Substituting the steady-state solution of Eq.~(\ref{eqatea}) into Eq.~(\ref{eqba}) and neglecting the term including $\Ateia$, we obtain
\begin{align}
\dt{\ba} &=   - \left(i \Dm + \frac{\Gm}{2} \right) \ba  \notag \\
&-  \frac{g^2 n^{\hspace{0.1mm}}_{\mathrm{TM}}}{-i\Dte + \left( \Gte /2 \right) } \ba  - \sqrt{\km} \Bia (t). \label{eqba2} \tag{S10}
\end{align}
At this stage, we consider the steady state of the Kittel mode $\ba$. Then, the coupled equations can be solved thoroughly to obtain the expression of $\atec$ as
\begin{align*}
\atec &= i g \sqrt{n^{\hspace{0.1mm}}_{\mathrm{TM}}} \frac{1}{i\Dte - \left( \Gte /2 \right) } \frac{\sqrt{\km}}{i\Dm + \left( \Gm /2 \right) } \Bia,
\end{align*}
where we assume that the coupling is weak, i.e., $g^2 n_{\rm{TM}} \ll \Gtm \Gm$.
To quantify the scattered TE-mode photon flux, the input-output relation $\Ateoa = \sqrt{\kte} \atea$ is invoked.  With this we can express the scattered photon flux as
\begin{align}
\nteo &= g^2 \rte (\wtm-\wm) \rtm (\wtm) \rmag (\wm)\notag \\
&\hspace{50mm}\times \ntmi (\nmi+1). \label{CCW-TMin} \tag{S11}
\end{align}
with the densities of states defined by
\begin{align*}
\rte(\wte) &= \frac{\kte}{\Dtesq + (\Gte /2)^2}\\ \rtm(\wtm) &= \frac{\ktm}{\Dtmsq + (\Gtm /2)^2}\\  \rmag(\wm) &= \frac{\km}{\Dmsq + (\Gm /2)^2}.\notag
\end{align*}
Thus, considering the contribution of every WGM labeled by $i$ in an FSR, we derive the scattering strength for the TM-mode input as $I_{\mathrm{TM \rightarrow TE}}(\omega) \equiv \sum_i C_i \rho^{(i)}_{\mathrm{TE}}(\omega - \omega_{\mathrm{mag}}) \rho^{(i)}_{\mathrm{TM}}(\omega)$ with the mode indices  $i$ and coefficients $C_i$ which include the coupling coefficients, the Kittel mode's density of state, and the input photon flux.  A similar procedure in the case of the TE-mode input results in the expression of the scattered photon flux
\begin{align}
\ntmo &= g^2 \rte (\wte) \rtm (\wte+\wm) \rmag (\wm) \ntei \nmi. \label{CCW-TEin} \tag{S12}
\end{align}
and the scattering strength for the TM-mode input $I_{\mathrm{TE \rightarrow TM}} (\omega) \equiv \sum_i C_i \rho^{(i)}_{\mathrm{TE}}(\omega) \rho^{(i)}_{\mathrm{TM}}(\omega + \omega_{\mathrm{mag}})$.

\section{Supplement C: Conversion efficiency and coupling constant}

\subsection{Estimation of the conversion efficiency and the coupling constant}

Regarding the optomagnonic system as a microwave-to-optical photon quantum converter, the estimation and the prospects for further improvement of the conversion efficiency are of crucial importance.  To evaluate the conversion efficiency $\nteo/\nmi$, we focus on the peak height of the sideband obtained in the experimental setup shown in Fig.~{\ref{FigB1}}(a).  A red-sideband spectrum for the TM-mode input is shown in Fig.~{\ref{FigB1}}(b) which is taken by the spectrum analyzer with the resolution bandwidth of 1~Hz.  Since the experiment is done with heterodyne measurement, the squared value of the beat note $\eta n_{\mathrm{LO}}n_s$ is observed at the spectrum analyzer where $\eta$ represents the coefficient accompanied to the conversion from optical power to the voltage at the photodetector, $n_{\mathrm{LO}}$ the photon flux of the LO light and $n_{s}$ that of the sideband signal.  The noise floor observed by the spectrum analyzer is the squared value of the photon shot-noise power accompanied to the LO light, namely, $\eta(\sqrt{2n_{\mathrm{LO}}})^2 = 2 \eta n_{\mathrm{LO}}$.  Therefore the signal-to-noise ratio is given in a simple form as ${n_s}/2$.  The shot-noise level at the spectrum analyzer is calibrated separately.  As in Fig.~{\ref{FigB1}}(b), the signal-to-noise ratio is 69~dB, which directly tells us the generated sideband photon number of $8.1\times10^6$.  Finally, as we know that the photon flux of the input microwaves is $2.2\times10^{20}$~/s for 0~dBm input and microwaves are critically coupled to the Kittel mode, the conversion efficiency can be evaluated as $7\times10^{-14}$ for the current experimental conditions.  

It is then possible to evaluate the value of optomagnonical coupling coefficient $g^{(\rm{exp})}$ using Eq.~(\ref{CCW-TMin}).  With the photon flux of th input optical field $\ntmi = 3\times10^{15}$~/s at 0.3~mW, the frequency difference between the TE and TM WGMs 50~GHz, and typical intrinsic Q factors of WGMs $1 \times 10^5$, $g^{(\rm{exp})}$ is calculated to be $2\pi\times5$~Hz.  Here we use the detuning of the input laser being 3~GHz lower than the input TM WGM, which maximizes the sideband strength in the experiment we performed.  External couplings are set to be 0.4~GHz for each WGMs regarding the transmission, and critical coupling for the Kittel mode.  The experimentally obtained value of the coupling $g^{(\rm{exp})}$ is consistent with the expected value $g^{(\rm{theory})}$.
A little deviation from the theoretical value may result from the imprecise choice of the parameters in $g^{(\rm{exp})}$, such as the cavity decay rates $\Gte$ and $\Gtm$.

\subsection{Prospect for improving the conversion efficiency}

In order to further improve the conversion efficiency, first we can change the size and shape of the WGM resonator so that the frequency difference between the TE and TM WGMs approaches the Kittel mode frequency.  In other words, it will be greatly beneficial to make the frequency matching condition $\wtm - \wte - \wm =0$ for microwave and optical fields involved be supported by the WGMs and the Kittel mode, $\Wtm - \Wte - \Wm =0$ (see Table~S1).  If this is done, the conversion efficiency will be $7\,000$ times larger.  Second, by properly polishing and chemically processing the WGM resonator one can get higher Q factor approaching the value limited by the optical absorption, $\sim 3\times10^6$ drived from the absorption coefficient $\alpha = 0.03$~/cm at room temperature \cite{Wood}.  This gives further improvement of the conversion efficiency by a factor of $3\,500$.  Third, the squared coupling coefficient $g^2$ is inversely proportional to the sample volume as can be seen in Eq.~(\ref{gth}).  Therefore, the non-spherical WGM resonator like disks and ellipsoids are of interest in this perspective.  For instance, a 1~$\mu$m-thick and 2.5~mm-diameter disk, which is expected to have a frequency difference between TE and TM WGMs of around 10 GHz according to the finite element method, reduces the volume by a factor of 90, and thus the conversion efficiency gains that factor.  All these enhancements together with optical input power of 20~mW will give the improved conversion efficiency of $3\times10^{-2}$.  With this value, the microwave-to-optical photon conversion enables various experiments, such as the optical control and readout of the quantum state of the superconducting qubit in the postselective way.

To reach unity photon conversion efficiency, additional ingredients should join the system.
Aside from the ways described above, a possibility for improvement is the reduction of the decay rate of the Kittel mode which becomes as small as 500 kHz at 1 K \cite{Tabuchi1}.  Other transparent materials with a large Verdet constant may further improve the efficiency to promisingly achieve a unity efficiency.


\end{document}